\documentclass[prd,preprint,eqsecnum,superscriptaddress,showpacs,
letterpaper]{revtex4}

\usepackage{color}
\usepackage{times}
\usepackage{graphicx}
\usepackage{fancyhdr}
\usepackage{float}

\renewcommand{\today}{\number\day\space\ifcase\month\or
  January\or February\or March\or April\or May\or June\or
  July\or August\or September\or October\or November\or December\fi
  \space\number\year}

\def\be{\begin{equation}}
\def\ee{\end{equation}}
\def\etc{{\it etc.}}

\def\etal{{\it et al.}}

\begin{document}

\title{Cumulative analysis of the association between the data of 
the Gravitational Wave detectors NAUTILUS and EXPLORER and the 
Gamma Ray Bursts detected by BATSE and BeppoSAX}

\begin{abstract}
\vspace*{0.2in}
The statistical association between the output of the Gravitational Wave
(GW) detectors EXPLORER and NAUTILUS and a list of Gamma Ray Bursts (GRBs)
detected by the satellite experiments BATSE and BeppoSAX has been analyzed
using cumulative algorithms. GW detector data collected between 1991 and 
1999 have been searched for an energy excess in a 10 s interval around the 
GRB flux peak times. The cumulative analysis of the data relative to a large 
number of GRBs (387) allows to push the upper bound for the corresponding 
GW burst amplitude down to $h = 2.5 \times 10^{-19}$.
\end{abstract}

\date[\relax]{Dated: \today }

\newcommand*{\RU}{INFN, Sezione di Roma, P.le Aldo Moro 2, I-00185, Roma, Italy}
\affiliation{\RU}
\newcommand*{\LF}{INFN, Laboratori Nazionali di Frascati, via Enrico Fermi 40,
I-00044, Frascati (Roma), Italy}
\affiliation{\LF}
\newcommand*{\TV}{Dipartimento di Fisica dell'Universit\`a ``Tor Vergata", 
 via della Ricerca Scientifica 1, I-00133, Roma, Italy}
\affiliation{\TV}
\newcommand*{\RD}{INFN, Sezione di Roma II, via della Ricerca Scientifica 1,
I-00133, Roma, Italy}
\affiliation{\RD}
\newcommand*{\AQ}{Dipartimento di Fisica dell'Universit\`a, via Vetoio (Coppito
1), I-67010, Coppito (L'Aquila), Italy}
\affiliation{\AQ}
\newcommand*{\GS}{INFN, Laboratori Nazionali del Gran Sasso, S.S. 17 bis km
18.910, I-67010, Assergi (L'Aquila), Italy}
\affiliation{\GS}
\newcommand*{\SP}{Dipartimento di Fisica dell'Universit\`a ``La Sapienza", 
P.le Aldo Moro 2, I-00185, Roma, Italy}
\affiliation{\SP}
\newcommand*{\FE}{Dipartimento di Fisica dell'Universit\`a, via Paradiso 12, 
I-44100, Ferrara, Italy}
\affiliation{\FE}
\newcommand*{\AS}{CNR, Istituto di Astrofisica Spaziale e Fisica Cosmica, 
via Piero Gobetti 101, I-40129, Bologna, Italy}
\affiliation{\AS}
\newcommand*{\SA}{ISA ``A.Venturi", Modena, Italy}
\affiliation{\SA}
\newcommand*{\FS}{CNR, Istituto di Fisica dello Spazio Interplanetario, 
via del Fosso del Cavaliere 100, I-00133, Roma, Italy}
\affiliation{\FS}
\author{P.~Astone}\affiliation{\RU}
\author{D.~Babusci}\affiliation{\LF}
\author{M.~Bassan}\affiliation{\TV}\affiliation{\RD}
\author{P.~Carelli}\affiliation{\AQ}\affiliation{\RD}
\author{E.~Coccia}\affiliation{\TV}\affiliation{\RD}\affiliation{\GS}
\author{C.~Cosmelli}\affiliation{\SP}\affiliation{\RU}
\author{S.~D'Antonio}\affiliation{\RD}
\author{V.~Fafone}\affiliation{\LF}
\author{F.~Frontera}\affiliation{\FE}\affiliation{\AS}
\author{G.~Giordano}\affiliation{\LF}
\author{C.~Guidorzi}\affiliation{\FE}
\author{A.~Marini}\affiliation{\LF}
\author{Y.~Minenkov}\affiliation{\TV}\affiliation{\RD}
\author{I.~Modena}\affiliation{\TV}\affiliation{\RD}
\author{G.~Modestino}\affiliation{\LF}
\author{A.~Moleti}\affiliation{\TV}\affiliation{\RD}
\author{E.~Montanari}\affiliation{\FE}\affiliation{\SA}
\author{G.~V.~Pallottino}\affiliation{\SP}\affiliation{\RU}
\author{G.~Pizzella}\affiliation{\TV}\affiliation{\LF}
\author{L.~Quintieri}\affiliation{\LF}
\author{A.~Rocchi}\affiliation{\TV}\affiliation{\LF}
\author{F.~Ronga}\affiliation{\LF}
\author{L.~Sperandio}\affiliation{\LF}
\author{R.~Terenzi}\affiliation{\FS}\affiliation{\RD}
\author{G.~Torrioli}\affiliation{\SP}
\author{M.~Visco}\affiliation{\FS}\affiliation{\RD}

\maketitle


\section{Introduction}\label{sec:intro}
Since 1991, almost 3000 Gamma Ray Bursts (GRBs) have been detected by the 
satellite experiments BATSE \cite{ref:fish,ref:briggs} and BeppoSAX
\cite{ref:boella,ref:costa}. The large database \cite{ref:pacie,ref:guido,
ref:flux} 
now available includes information about the GRB arrival time, duration, 
intensity in some frequency bands, sky position of the source, and (for a 
small GRB subset) redshift. The observation of a large number of GRBs, 
which are likely associated to catastrophic events capable of producing 
large GW signals, has given the possibility of systematic analysis of the 
GW detector data around the GRB arrival times. This is very important, 
because GW data analysis in association with GRBs can profit of a number 
of useful information (GRB time, source position, intensity \etc) and both 
positive and negative results could 
be given a direct astrophysical interpretation. Cumulative data analysis 
techniques have been developed to detect a statistically significant 
association between GW signals and GRBs 
\cite{ref:finn,ref:mode00,ref:murphy,ref:vinogra,ref:mode02}. Using for the 
first time 
a cross-correlation method applied to the data of two GW detectors, EXPLORER 
and NAUTILUS, experimental upper limits were determined for the amplitude of 
the GW bursts associated with GRBs \cite{ref:cc01}. Analyzing the data for 
47 GRBs detected by BeppoSAX, the presence of GW pulses of amplitude 
$h \ge 1.2 \times 10^{-18}$ was excluded with 95 \% probability, within the 
time window of $\pm$ 400 s. Within the time window of $\pm$ 5 s, the upper 
limit was improved to $h = 6.5 \times 10^{-19}$.

Searching for an association between the two emissions, the main difficulty 
arises from the theoretical uncertainty in the delay between the GRB and GW 
arrival times. All the theoretical models presently available 
\cite{ref:rees,ref:piran,ref:mesza,ref:zhang,ref:deru,ref:koba,ref:vput}, 
and the interpretation of experimental observations of GRB characteristics 
\cite{ref:front00,ref:front03,ref:sari}, 
foresee that the GRB generation can happen during different phases of 
catastrophic events involving binary systems or massive stars. During some of 
these phases, the GW emission could happen at the same time of the GRB one.
Thus, it is interesting to apply cumulative techniques making the restrictive 
hypothesis of simultaneity of the GRB and GW emissions. Implicitly making 
this hypothesis, several analyses have been performed 
\cite{ref:mode97,ref:vulc98,ref:tric}. In \cite{ref:tric} an upper limit 
of $h = 1.5 \times 10^{-18}$ on the average amplitude of GW associated to 
GRBs was obtained with the resonant bar detector AURIGA, using 120 GRBs and 
an integration time window of 10 s. 

According to the present knowledge of the GRB physics, at distance of
1 Gpc, GW burst signals of the order of $h \sim 10^{-22}$ or smaller are
expected in association with GRBs. At the time the data used here were 
taken, EXPLORER and NAUTILUS were probably the most sensitive GW detectors, 
having a sensitivity for 1 ms duration GW burst with signal-to-noise ratio 
equal to unity of about $10^{-18}$ in $h$, further improved in the following 
years \cite{ref:coi01}. Thus we expect a null result, which, however, can 
be used to set upper limits to the GW flux. The present limits need to be 
significantly improved to get useful constraints on current GRB theoretical 
models. Recently, the large interferometric GW detectors are beginning to 
come into operation, and in particular LIGO is reaching a sensitivity that 
allows to start looking at correlation with GRBs \cite{ref:ligogrb}. 

In section \ref{sec:datam} the data and the cumulative algorithms used in 
this work will be described \cite{ref:ama03}. The results will be shown and 
discussed in section \ref{sec:resdi}.

\section{Data and Method}\label{sec:datam}
\begin{figure}[!ht]
\centering
\includegraphics[height=4.2in,width=5.1in]{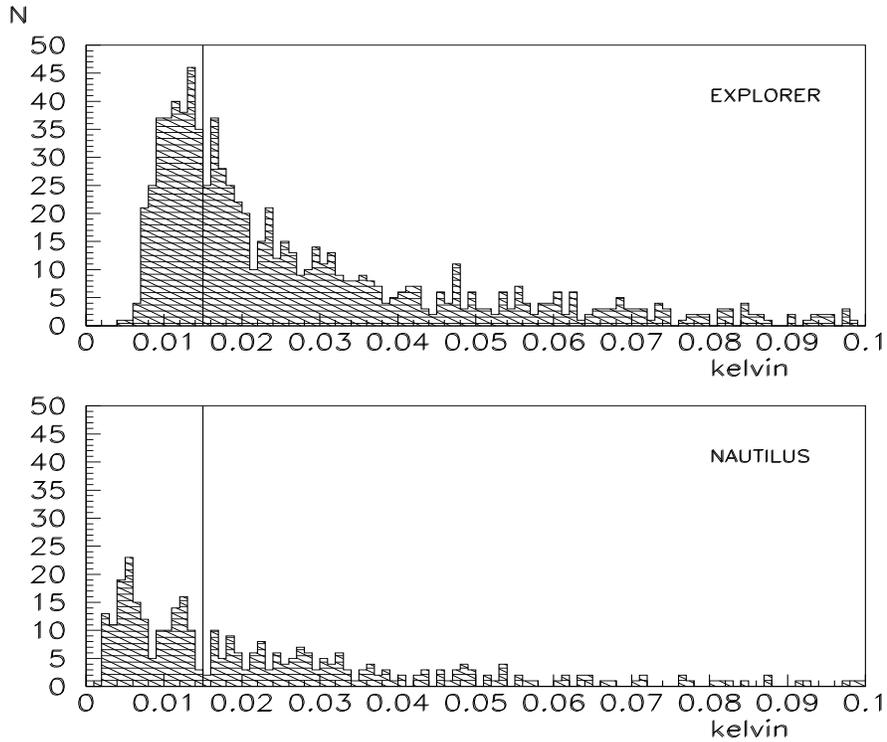}
\caption{Histograms of the effective temperature of the ``minimum"
mode of the Wiener filtered data computed in the 1150 time intervals
of 30 min around each GRB time.}\label{fig:distri1}
\end{figure}

The ROG Collaboration operates two resonant bar detectors: EXPLORER since 
1990 at the CERN laboratory and NAUTILUS since 1995 at the INFN laboratory 
in Frascati. The two detectors, oriented nearly parallel, are very similar. 
They consist of massive cylindrical bars 3 m long made of high quality factor 
aluminum alloy 5056. The GW excites the first longitudinal mode of the bar 
which is cooled to liquid helium temperature to reduce the thermal noise. 
To measure the bar strain induced by a GW, a secondary mechanical oscillator 
tuned to the antenna mode is mounted on one bar face (as a consequence we 
have two resonant modes) and a sensor measures the displacement between the 
secondary oscillator and the bar face. The frequencies of these resonant modes 
varied slightly during the years, remaining for both antennas in the range 
900-940 Hz. The data considered in the present analysis are sampled with  
a sampling time of 0.2908 s and processed with an adaptive Wiener filter 
\cite{ref:ast94}. The Wiener filtered data represent the energy innovation 
(expressed in kelvin) of each of the two modes. For each data sample, the 
minimum energy between the two modes is taken, obtaining the ``minimum" mode 
time series, $E(t)$, which is the one used in this analysis. The probability 
distribution of $E(t)$ is
\be
f(E) \propto  \frac{1}{T_{\rm eff}}\,{\rm e}^{- \frac{E}{T_{\rm eff}}}\,,
\label{normal}
\ee
where $T_{\rm eff}$, called {\em effective temperature} and expressed in 
kelvin units, gives an estimate of the noise. In our analysis data stretches 
of 30 min duration were considered, centered at the arrival times of the GRBs. 
In Fig. \ref{fig:distri1}, the distribution of $T_{\rm eff}$ is shown for 
1150 data stretches selected for the analysis. The upper histogram corresponds 
to the EXPLORER data, the second one to the NAUTILUS data. 

As regards the quality of the GW data, in order to improve the
sensitivity of the analysis, we only consider the data stretches with
effective temperature lower than 15 mK. In addition, we request that
the ratio between the standard deviation and the average of each GW
data stretch (this ratio is expected to be unity for an exponential
distribution) lies between 0.8 and 1.5. These selection criteria restrict 
the data set to 387 GRBs. As GRB arrival time, we define the time of the 
flux peak on the 1024 ms trigger time scale extracted from the 
{\it Flux and Fluence Table} of {\it BATSE Current GRB Catalog} 
\cite{ref:flux}, 
while for BeppoSAX the GRB peak time is given by the time of the peak flux 
on a 1 s integration time. The GRB data also provide the angular position 
of each source, which is an important parameter, because the sensitivity of 
a cylindrical bar GW detector is strongly dependent on the angle $\theta$ 
between the propagation direction of the wave and the axis of the cylinder. 

\begin{figure}
\centering
\includegraphics[height=3in,width=4in]{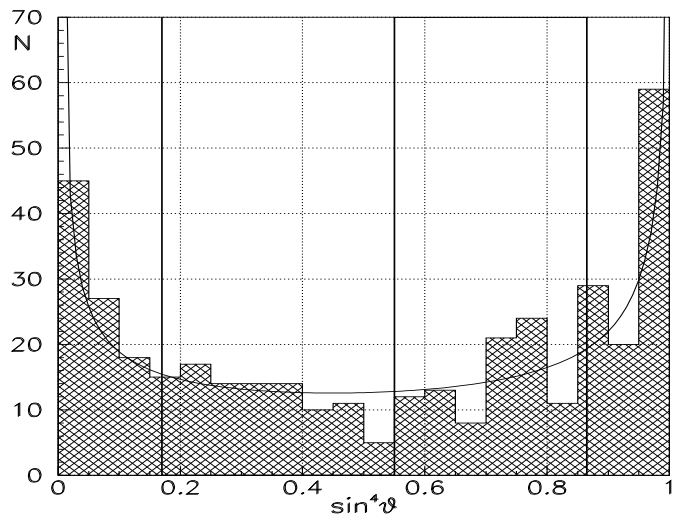}
\caption{Experimental histogram of $\sin^4 \theta$ for the 387 GRB
in the selected time intervals of 30 min around each GRB time (net area).
The distribution is compared to theoretical isotropic distribution
(solid line). The four regions of increasing $\sin^4 \theta$, 
separated by vertical lines, correspond to the data subsets separately 
analyzed to look for a correlation with $\sin^4 \theta$.}\label{fig:distri2}
\end{figure}

The histogram of Fig. \ref{fig:distri2} shows the distribution of 
$\sin^4 \theta$ for the 387 GRBs corresponding to the selected data stretches 
with $T_{\rm eff} \leq 15$ mK. The distribution has been compared to the 
theoretical distribution expected for isotropic sources by the Kolmogorov test 
\cite{ref:kolmo}. 
The result of the test indicates a compatibility more than 0.9 in terms of 
probability. It means that in the present analysis there is no privileged 
direction. As we can note, the data sample is large enough to look for a 
statistical correlation between the presence of a GW energy excess at zero 
delay and the value of $\sin^4 \theta$. For this, the data set is divided 
into four equally populated ranges of $\sin^4 \theta$, as indicated in 
Fig. \ref{fig:distri2} by the vertical 
lines, then these regions will be separately analyzed.

In the present work we use two algorithms, both based on coherent averages 
performed over the selected GW data stretches synchronized using the GRB 
flux peak time as a common reference in order to show a possible energy
excess at zero-delay time within an integration time of 10 s \cite{ref:integ}.

The first algorithm computes the average of the data stretches
corresponding to each GRB: we construct a new data stretch where at each 
time there is the average of the values, at that same time, of all the
measured data stretches. The averaged energy at zero-delay is the
measured physical quantity to be compared with the distribution of the 
same averages taken at all the other times, constituting the background. 

The second algorithm, which is a new one for this kind of analysis, 
differs from the first one since it uses the median of the data instead of 
the average. This is a robust way to detect the effect of many small 
synchronized contributions rather than that of a single or of a few very 
large signals.
Indeed, it is easy to understand that a few intense spikes increase
the variance of the average much more than that of the median.
This is important also because the noise distribution of GW 
detectors data is affected by significant non-gaussian tails,
thus the occurrence of intense spurious noise spikes is not as
unfrequent as it would be for an ideal detector with gaussian noise.

\section{Results and Discussion}\label{sec:resdi}
In this work cumulative algorithms were used, searching
for an energy excess above the background of the GW data at the GRB
arrival time. Thus the results of this analysis, in terms of signal
detected or upper limits, represent the average GW flux associated to
each GRB and released simultaneously to the gamma emission, within a
given time interval, telling nothing about the possibility of a much
earlier and time-scattered GW emission. The analysis of a much larger time
interval (30 min around the GRB time), which is performed in this work,
has the purpose of estimating the background statistical distribution of
the physical quantity that, computed at zero delay, is assumed to be the
indicator of correlation with GRBs.
\begin{figure}[!h]
\centering
\includegraphics[height=4in,width=4.5in]{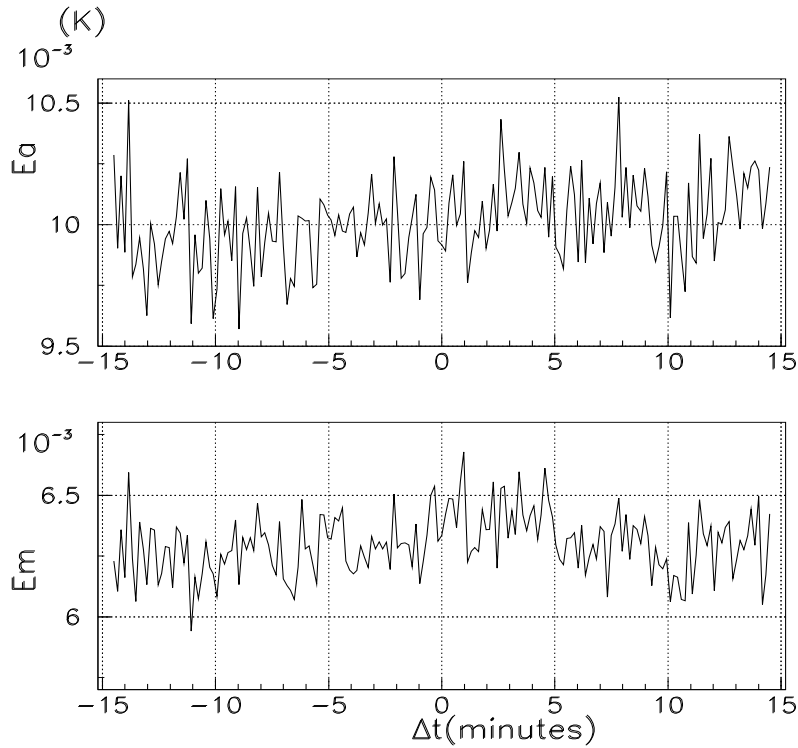}
\caption{Cumulative average ($E_a$) and cumulative median ($E_m$) of 
the GW detector energy as a function of the GW-GRB delay.}\label{fig:distri3}
\end{figure}

In Fig. \ref{fig:distri3} the result of the application of the average
algorithm is shown. The averaged GW detector energy innovation is
plotted as a function of time, relative to the GRB flux peak time.
In the same figure, the result of the application of the second
algorithm is also reported. In this case, for each 10 s interval,
the median of the distribution of the GW detector energy innovation
measured in that interval is shown, as a function of the GW-GRB delay. 

\begin{figure}[!h]
\centering
\includegraphics[height=3.5in,width=4.5in]{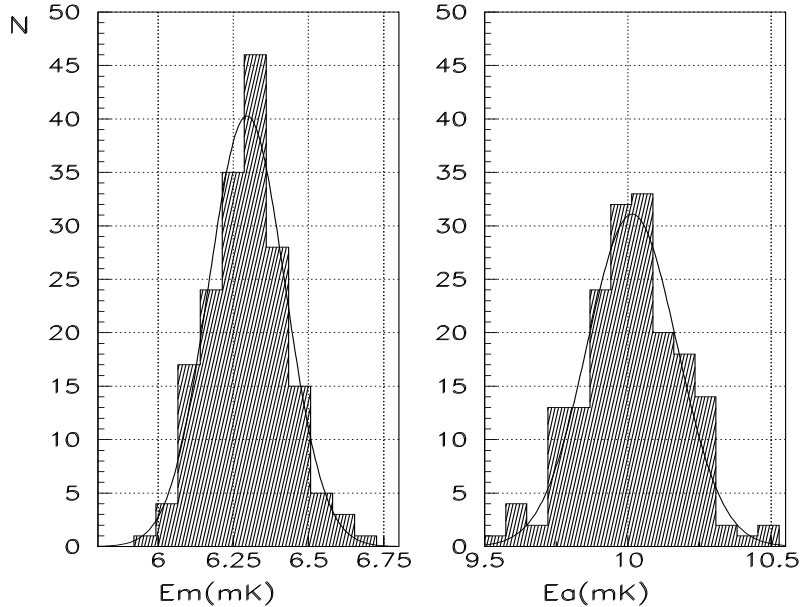}
\caption{Distributions of the median and of the average of the GW
detector energy value (see Fig. \ref{fig:distri3}) and gaussian fits.}
\label{fig:distri4}
\end{figure}

From the average and median time series shown in Fig. \ref{fig:distri3}, 
$E_a(t)$ and $E_m(t)$, we consider the average and median value at 
zero delay, $E_a(0)$ and $E_m(0)$, and compute the time averages 
$<E_a>$ and $<E_m>$, and the standard deviations $\sigma_a$ and 
$\sigma_m$ of the values at all other times, finding:
\begin{eqnarray}
{\rm average:} & ~~E_a(0) = 9.91 ~{\rm mK}, ~~~~<E_a>\,=\,10.01 ~{\rm mK}, &
~~~~\sigma_a=0.17 ~{\rm mK}; \nonumber \\[-2mm]
 & & \\[-2mm]
{\rm median:}  & ~E_m(0)=6.33 ~{\rm mK}, ~~~~<E_m>\,=\,6.30 ~{\rm mK}, &
~~~~\sigma_m=0.13 ~{\rm mK}. \nonumber
\end{eqnarray}
The distributions show a good fit with the gaussian curves. For example, 
the agreement is shown in Fig. \ref{fig:distri4} for the distributions
relative to the Fig. \ref{fig:distri3}.

\begin{figure}
\centering
\includegraphics[height=3.5in,width=4.5in]{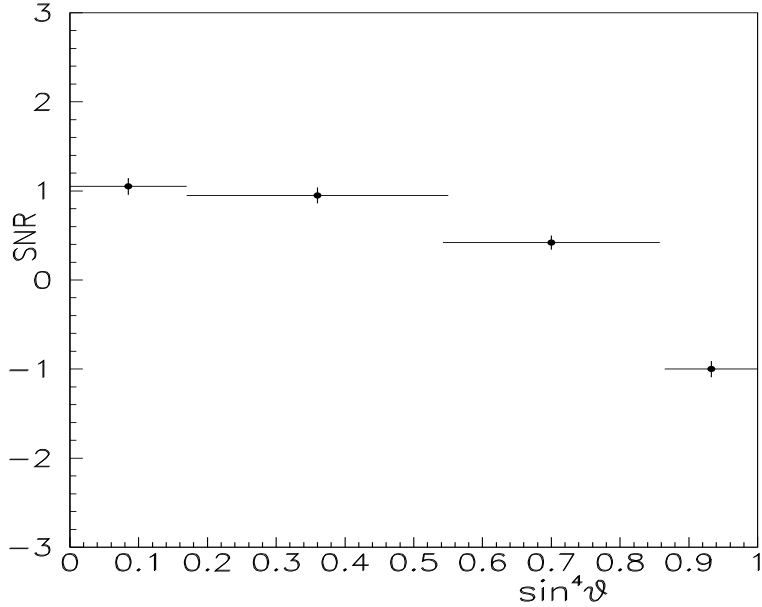}
\caption{SNR of the excess at zero delay of the GW-median, as a function
of $\sin^4 \theta$.}\label{fig:distri6}
\end{figure}

With respect to the dependence of the observed energy value on 
$\sin^4 \theta$, at zero-delay, the source direction information was
used by separately analyzing the GRBs whose average $\sin^4 \theta$
factor is within a given interval during the 30 min interval. The result 
of this analysis, shown in Fig. \ref{fig:distri6}, was obtained applying 
the median algorithm to four subsets of GRBs, whose possible GW sources would 
be increasingly well-placed in the sky relative to the antenna axis, as 
expressed by their average value of $\sin^4 \theta$. The quantity plotted 
in Fig. \ref{fig:distri6} is, for each subset, the signal-to-noise ratio, 
defined as:
\be
{\rm SNR} \equiv {\frac{E_m(0)- <E_m>}{\sigma}}\;,
\ee
where $E_m(0)$ is the value of the median at zero delay, $<E_m>$ and 
$\sigma$ are the average and standard deviation of all the values at 
non-zero delay in the cumulative median time series (see 
Fig. \ref{fig:distri3}). 
The vertical bars indicate the uncertainty in SNR as deduced from the ones 
in $<E_m>$ and $\sigma$. No clear correlation (i.e. with SNR $>$ 1) is 
visible in the data with the average value of $\sin^{4} \theta$.

\section{Upper limit evaluation}\label{sec:uplim}
Fig. \ref{fig:distri4} shows that both the average and median distributions 
are close to normal. This allows us to represent the sensitivity of the 
experiment as a function of $h$ and to evaluate an upper limit, using the 
same approach followed in our previous GRB-GW coincidence analysis 
\cite{ref:cc01}, based on the likelihood rescaled to its value for background 
alone ($R$ function, called also \emph{relative belief update function} 
\cite{ref:gda}). In fact, in the Bayesian approach we are implicitly following, 
the likelihood has the role of modifying our knowledge according to the scheme 
\emph{posterior} $\propto~R ~\times$ \emph{prior}. In presence of a signal with 
energy $E_s$ we expect to measure an energy $E_b$ larger by some quantity 
$\delta$ with respect to the case of no signal, that is:
\be
E_b = E_n + \delta,
\ee
where $E_n$ is due to noise. We indicate the measurement at zero time
delay with $E_0$. Thus the expected normal distribution is
\be
f(E_0|\delta) \sim e^{-(E_0-(E_n+\delta))^2/2\sigma^2},
\ee
where $\sigma$ is the experimental standard deviation. We find the 
\emph{relative belief updating ratio R}
\be 
R(\delta) = \frac{f(E_0|\delta)}{f(E_0|\delta=0)} = 
e^{-(\delta^2-2E_0\delta+2E_n\delta)/2\sigma^2}.
\ee

Using the quantities defined in the previous section, we can compute the 
functions $R_a(\delta_a)$ and $R_m(\delta_m)$, in the case of the average 
and median algorithm respectively, and so we obtain an \emph{upper limit}, 
or, better, an \emph{upper sensitivity bound} on the value of $\delta_a$ 
and $\delta_m$. If we take conventionally $R(\delta)=0.05$, we determine
\be
\delta_a(5\%)\sim 0.33 ~\rm{mK},\qquad \delta_m(5\%)\sim 0.35 ~\rm{mK}.
\ee
In order to find the relation between the increase $\delta_a$ and the 
corresponding value $E_s$ of the signals that would generate it, we have 
to take into account that, as we discussed in section \ref{sec:datam}, 
we take time averages of 10 s, and this leads to a loss in sensitivity, 
since the signal due to a GW burst would usually be shorter than 10 s. 
We evaluate this sensitivity loss in a factor 3.

In the case of the median algorithm, a further factor comes out: in the 
hypothesis of $E_s$ much smaller that $E_n$, the distribution of $E_n+E_s$ 
remains exponential (as it was for $E_n$) and so if the average energy 
increases by $E_s$ the median value increases by $E_s \ln 2$.

The energies $E_s^{a,m}$ corresponding to the values $\delta_{a,m}(5\%)$ are 
then
\be
E_s^a \sim 1 ~{\rm mK}, \qquad E_s^m \sim 1.5 ~{\rm mK}.
\label{eqes}
\ee

\begin{figure}[t]
\centering
\includegraphics[height=3.5in,width=4.5in]{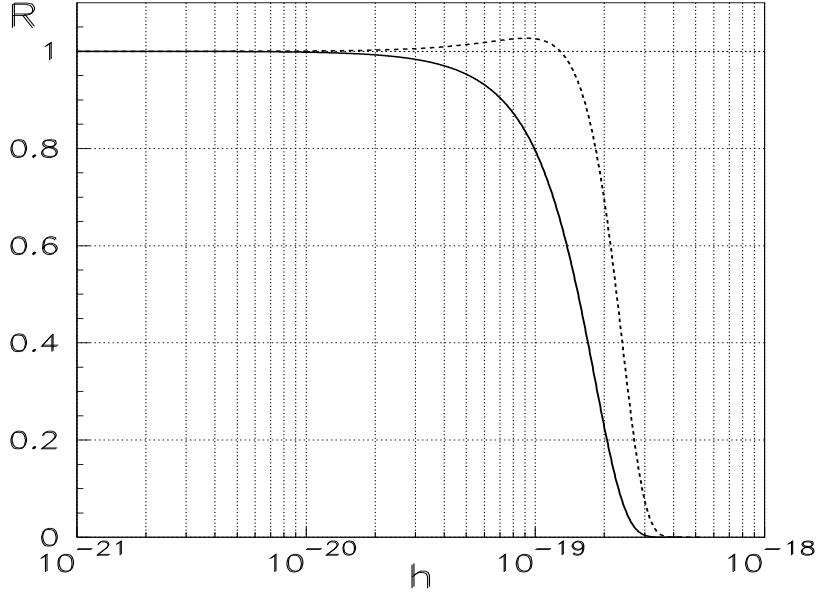}
\caption{\emph{Relative belief updating ratio} as a function of $h$,
using the average (solid line) and median (dashed line) algorithms.}
\label{fig:distri7}
\end{figure}

The signal energy $E_s$ is determined by the value of the Fourier transform 
$H(f)$ of the GW in the detector frequency band; computation of the GW burst 
amplitude $h$ requires a model for the signal shape. A conventionally chosen 
shape is a featureless pulse lasting a time $\tau_g$ and giving a constant 
Fourier spectrum over a frequency band equal to $1/\tau_g$. Assuming the 
detector band within this range, for optimal orientation one has:
\be
h = \frac{H}{\tau_g}\,=\,\frac{1}{\tau_g}\,\frac{L}{v_s^2}\,
\sqrt{\frac{k E_s}{M}},
\label{aus}
\ee
where $v_s = 5.4$ km$\;{\rm s}^{-1}$ is the sound velocity in aluminum, $L$ and $M$ 
are the length and the mass of the bar, respectively. We conventionally assume 
a GW burst duration $\tau_g$ = 1 ms, so the $E_s$ values of Eqn. \ref{eqes} 
correspond to two quite close values for the sensitivity bound in $h$:
\be
h_a \sim 2.5 \times 10^{-19}, \qquad h_m \sim 3.1 \times 10^{-19}.
\ee
The behaviour of the \emph{relative belief updating ratio R} as a function 
of $h$ is given in Fig. \ref{fig:distri7}, in both the average and median 
cases. We notice that in both cases, $R \simeq$ 1 in the region with
$h \leq 2 \times 10^{-20}$: this means that the detectors were not sensitive 
enough to appreciate such small amplitudes, and hence nothing can be learned 
from the experiment in that region of $h$.

\section{Conclusions}\label{sec:conclu}
A large sample of GRBs (387) was used, to search for an association
between the GW detector data and GRBs at zero delay. No statistically
significant excess was observed at zero delay, within the time resolution
of 10 s. We performed an analysis based on a Bayesian approach, obtaining 
an upper bound on the GW burst amplitude associated with GRB of 
$h \sim 2.5 \times 10^{-19}$. 

\begin{acknowledgments}
We thank Peter Saulson and other members of LSC for useful discussions 
and comments. 
\end{acknowledgments}

\end{document}